\documentclass[10pt,twocolumn,letterpaper]{article}

\usepackage{cvpr}
\usepackage{times}
\usepackage{epsfig}
\usepackage{graphicx}
\usepackage{amsmath}
\usepackage{amssymb}
\usepackage{subfigure}

\usepackage[breaklinks=true,bookmarks=false]{hyperref}

\cvprfinalcopy 

\def\cvprPaperID{****} 

\begin{document}

\title{Performance Comparison of Convolutional AutoEncoders, Generative Adversarial Networks and Super-Resolution for Image Compression}

\author{Zhengxue Cheng, Heming Sun, Masaru Takeuchi, and Jiro Katto\\
Department of Computer Science and Communications Engineering \\
Waseda University, Tokyo, Japan\\
{\tt\small zxcheng@asagi.waseda.jp, terrysun1989@akane.waseda.jp, masaru-t@aoni.waseda.jp, katto@waseda.jp}
}

\maketitle

\begin{abstract}
   Image compression has been investigated for many decades. Recently, deep learning approaches have achieved a great success in many computer vision tasks, and are gradually used in image compression. In this paper, we develop three overall compression architectures based on convolutional autoencoders (CAEs), generative adversarial networks (GANs) as well as super-resolution (SR), and present a comprehensive performance comparison. According to experimental results, CAEs achieve better coding efficiency than JPEG by extracting compact features. GANs show potential advantages on large compression ratio and high subjective quality reconstruction. Super-resolution achieves the best rate-distortion (RD) performance among them, which is comparable to BPG.
\end{abstract}

\vspace{-3mm}
\section{Introduction}

Image compression has been a fundamental and significant research topic in the field of image processing for several decades. Traditional image compression algorithms, such as JPEG~\cite{IEEEexample:JPEG}, JPEG2000~\cite{IEEEexample:JPEG2000} and BPG~\cite{IEEEexample:BPG}, rely on hand-crafted encoder/decoder (codec) block diagrams. They use fixed transforms, i.e. Discrete Cosine Transform (DCT) and Discrete Wavelet Transform, together with the quantization and the entropy coder to reduce spatial redundancy for natural scene images. However, they are not expected to be an optimal and flexible image coding solution for all types of image contents and formats.

Deep learning approaches has the potential to enhance the performance of image compression. Recently, several methods have been proposed using different neural networks. The works~\cite{IEEEexample:Theis}\cite{IEEEexample:Balle} proposed a differentiable approximation of the quantization and the entropy rate estimation for an end-to-end autoencoder. The work~\cite{IEEEexample:Toderici} used a recurrent network for compressing full-resolution images. Priming and spatially adaptive bit rate were further considered in~\cite{IEEEexample:Nick}. Generative adversarial networks (GANs) were used for image compression in~\cite{IEEEexample:MITgan} and ~\cite{IEEEexample:Waveone}, which achieved better performance than BPG. Neural networks based super-resolution methods achieve better quality than conventional interpolation methods, so it can be used as a post filter to enhance the compression performance. Deep learning based approaches not only achieve better coding efficiency, but also can adapt much quicker to new media contents and new media formats~\cite{IEEEexample:Theis}. Therefore, learned image compression is expected to be more efficient and more general.



In this paper, we propose three architectures using convolutional autoencoders (CAEs), GANs and super-resolution (SR) for lossy image compression, respectively. Moreover, we discuss their coding performance and present a comprehensive comparison. Experimental results demonstrate that CAEs achieve higher coding efficiency than JPEG due to the property of compact representation of autoencoders. GANs show potential advantages on large compression ratio and high subjective quality reconstruction. Super-resolution achieves the best rate-distortion (RD) performance among three methods.


\section{Three Image Compression Methods}

\subsection{Convolutional Autoencoders for Compression}


Generally, an autoencoder can be regarded as an encoder function, $y=f_{\theta}(x)$, and a decoder function, $\hat{x}=g_{\phi}(y)$, where $x$, $\hat{x}$, and $y$ are original images, reconstructed images, and compressed data, respectively. $\theta$ and $\phi$ are optimized parameters in the encoder and the decoder function.

\begin{figure}[tb]
	\centerline{\psfig{figure=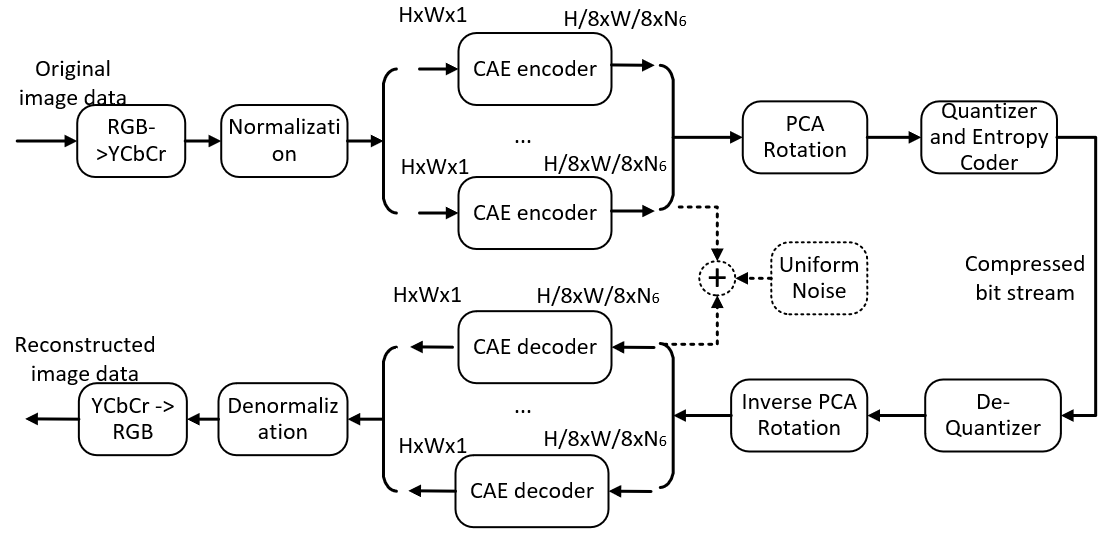,width=80.7mm} }
	\caption{Block diagram of CAE based image compression.}
	\label{fig:CAEoverall}
\end{figure}

\begin{figure}[tb]
	\centerline{\psfig{figure=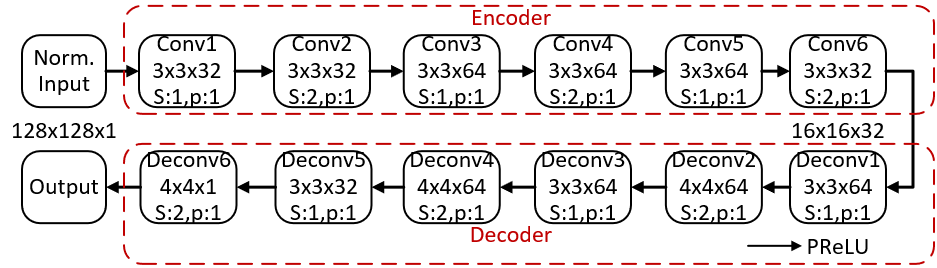,width=78.7mm} }
	\caption{The CAE network structure.}
	\label{fig:CAE}
\end{figure}

We propose a CAE network to replace conventional transforms, such as DCT and wavelet transform. The overall architecture is shown in Figure~\ref{fig:CAEoverall}. Consecutive downsampling operations destroy the quality of reconstructed images. Therefore, we use a pair of convolution/deconvolution filters for one upsampling or downsampling operation. The CAE network structure is shown in Figure~\ref{fig:CAE}. As for the activation function after each convolutional layer, we utilize the Parametric Rectified Linear Unit (PReLU) function, instead of ReLU, which is commonly used in related works, because we find that PReLU can improve the quality of reconstructed images compared to ReLU, especially with high bit rate. Inspired by the RD cost function in traditional codecs, the loss function is defined as
\begin{equation}
\begin{aligned}
J(\theta, \phi; x) &= ||x-\hat{x}||^{2}+\lambda\cdot||y||^{2}\\
&= ||x-g_{\phi}(f_{\theta}(x)+\mu)||^{2}+\lambda\cdot||f_{\theta}(x)||^{2}\\
\end{aligned}
\end{equation}
where $||x-\hat{x}||^{2}$ denotes the mean square error (MSE) distortion between original images $x$ and reconstructed images $\hat{x}$. $\mu$ denotes uniform noises. $\lambda$ controls the tradeoff between the rate and the distortion. $||f_{\theta}(x)||^{2}$ denotes the amplitude of compressed data $y$, which reflects the number of bits used to encode compressed data. We use a subset of ImageNet database~\cite{IEEEexample:ImageNet} consisting of 5500 images to train the CAE network. We used the Adam optimizer~\cite{IEEEexample:adam} and a batch size of 16 to train the model up to $8\times10^{5}$ iterations. The learning rate was kept at a fixed value of $0.0001$, and the momentum $\beta_{1}$ was set as $0.9$. Then we apply the principle component analysis (PCA), uniform quantization and the JPEG2000 entropy coder to generate a bit stream.

\subsection{Generative Adversarial Networks for Compression}

For the GAN based image compression, we add one convolutional layer to make the input size as $128\times128$, based on the architecture of DCGAN~\cite{IEEEexample:MITgan}\cite{IEEEexample:DCGAN}. The activation function is kept the same as DCGAN. Because DCGAN only includes the generator as the decoder function, we add an encoder function, which has the same structure as the discriminator. To implement the end-to-end training, the loss function of the generator is defined as
\begin{equation}
J_{G}(x) = ||x-\hat{x}||^{2}+\beta\sum_{i\in[0,4]}||D_{hi}(x) - D_{hi}(\hat{x})||^{2}
\end{equation}
where $||x-\hat{x}||^{2}$ denotes the MSE distortion between the original images $x$ and reconstructed images $\hat{x}$. Adding the discriminator network benefits the high quality reconstruction~\cite{IEEEexample:Waveone}, so we add the second distortion term in Eq.(2). $D_{hi}(x)$ and $D_{hi}(\hat{x})$ are the outputs of the $i$-th convolutional layer in discriminator network for inputs $x$ and $\hat{x}$, respectively. $\beta$ is set as $0.01$ in our experiments. The loss function of the discriminator is kept the same as DCGAN.

\begin{figure}[tb]
	\centerline{\psfig{figure=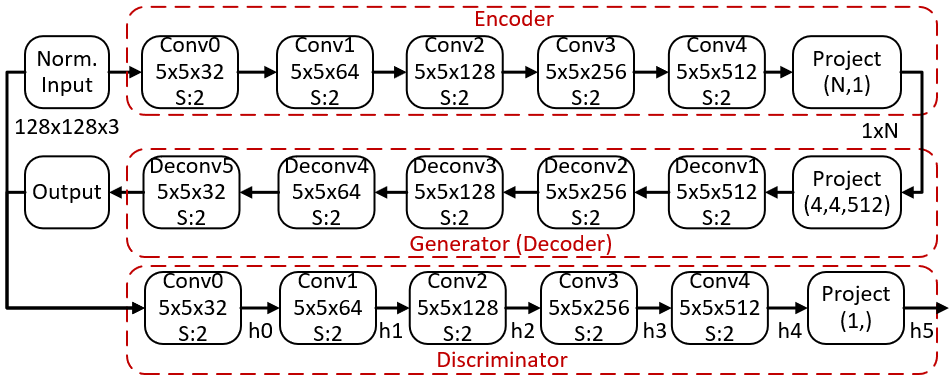,width=78.7mm} }
	\caption{The GAN structure.}
	\label{fig:GAN}
\end{figure}

We use the training set of the \emph{Workshop and Challenge on Learned Image Compression (CLIC)}. The Adam optimizer~\cite{IEEEexample:adam} with a batch size of 128 was used for training. The learning rate was kept at a fixed value of $0.0001$. The model is trained up to 25 epoches. The GAN structure is shown in Figure~\ref{fig:GAN}. The GAN based architecture has three differences from the CAE based architecture. First, the input has RGB components, so color space conversion from RGB to YCbCr is not applied. Second, we do not add uniform noises during the training process since GAN inherently reconstructs images from noises. Third, we use the range coder~\cite{IEEEexample:rangecoder}, instead of the JPEG2000 entropy coder.

\begin{figure}[tb]
	\centerline{\psfig{figure=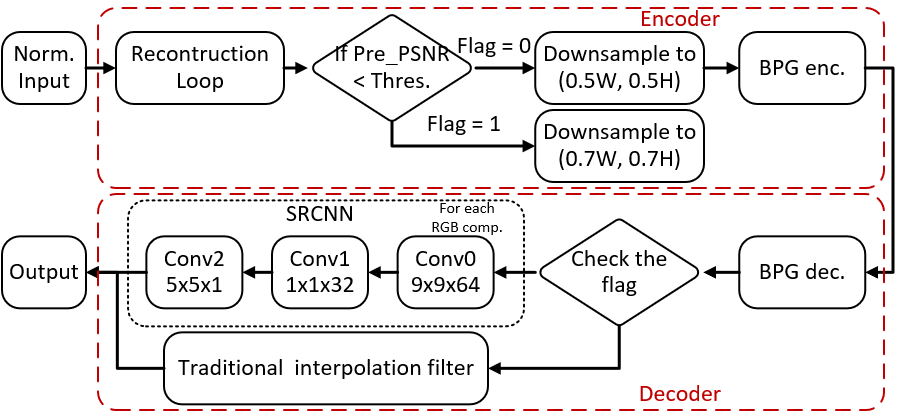,width=72.7mm} }
	\caption{Block diagram of super-resolution based compression.}
	\label{fig:SRCNN}
\end{figure}

\subsection{Super-Resolution for Compression}

Using super-resolution as a post filter is an intuitive method for compression. We present a SR based compression architecture in Figure~\ref{fig:SRCNN}. We use a SRCNN architecture in~\cite{IEEEexample:SRCNN} with three convolutional layers. The kernel sizes are set as 9, 1, 5, and the numbers of convolutional filters are set as 64, 32, 1. We retrain this SRCNN model with the scale of 2 using the \emph{CLIC} training dataset. The loss function and training parameters are kept the same as~\cite{IEEEexample:SRCNN}.

However, for images with complex textures or with small resolution, SR will become the bottleneck of high quality reconstruction. Thus, we propose an adaptive strategy by building a reconstruction loop in the encoder. This loop calculates the distortion only caused by SR, i.e. Pre\_PSNR in Figure~\ref{fig:SRCNN}. When Pre\_PSNR is larger than a pre-defined threshold, images are downsampled to (0.5W, 0.5H) and a SRCNN filter is conducted after decoding. Otherwise, images are downsampled to (0.7W, 0.7H) and a lanczos filter is alternatively applied for interpolation. The effect of adaptive strategy is listed in Table~\ref{tab.adaptive}. The threshold is set as $33.0$ dB in our experiments and about 30\% of images are selected to use SRCNN filters. For the \emph{CLIC} challenge, the entry \emph{Kattolab} uses adaptive SR-based architecture.



\begin{table}
\begin{center}
\begin{footnotesize}
\begin{tabular}{|l|l|l|l|}
\hline
\textbf{Case} & \textbf{PSNR (dB)}  & \textbf{MS-SSIM} & \textbf{Rate (bpp)} \\
\hline
qp=32, Non-adaptive & 29.418 &0.949 &0.151\\
qp=35, Adaptive & 30.002  &0.945 & 0.156\\
\hline
\end{tabular}
\end{footnotesize}
\end{center}
\caption{The effect of adaptive strategy for super-resolution.}
\label{tab.adaptive}
\end{table}

\vspace{-2mm}
\section{Performance Discussion and Comparison}

To measure the coding efficiency, the rate is measured by bit per pixel (bpp). PSNR (dB) and MS-SSIM are used to measure objective and subjective qualities, respectively.

\subsection{Discussion on CAE-based Compression}

\begin{figure}[tb]
\centering
\subfigure[]{
\label{Fig.sub.1}
\includegraphics[width=0.09\textwidth]{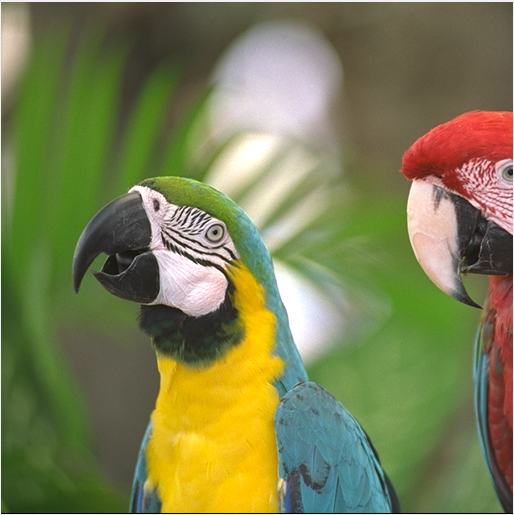}}
\subfigure[39.2dB, 1.31bpp]{
\label{Fig.sub.2}
\includegraphics[width=0.18\textwidth]{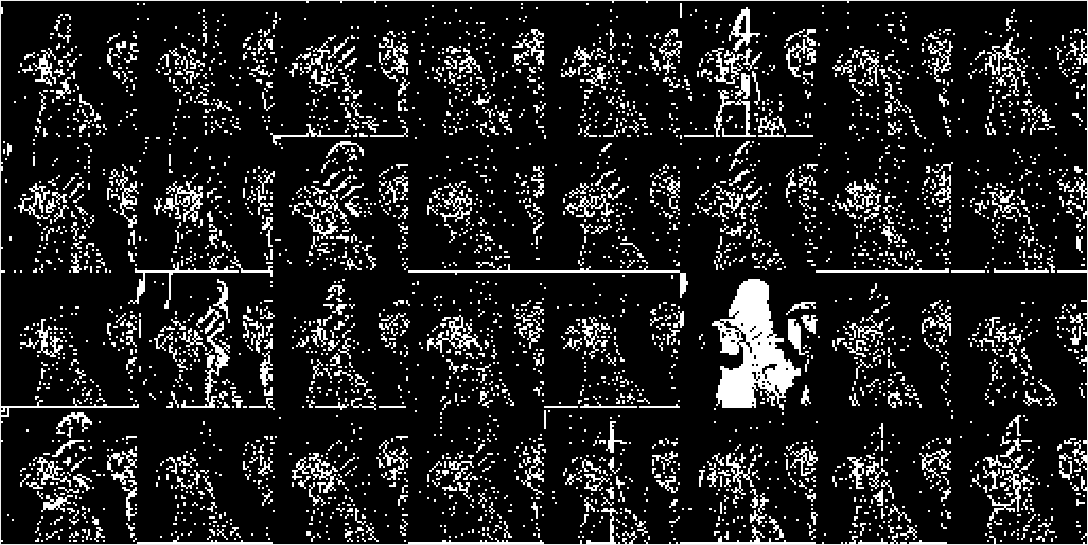}}
\subfigure[39.2dB, 0.99bpp]{
\label{Fig.sub.9}
\includegraphics[width=0.18\textwidth]{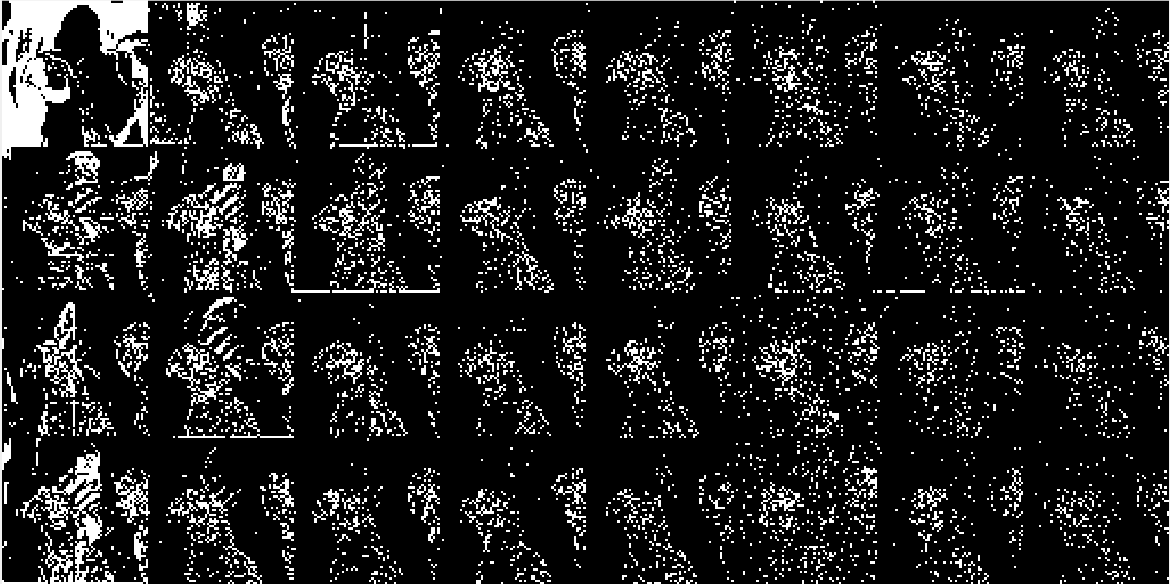}}
\caption{One image and corresponding 32 feature maps of Y-component generated by CAE, arranged in raster-scan order and rotated feature maps by PCA, arranged in vertical scan order.}
\label{fig:features}
\end{figure}

The feature maps generated by CAEs are not energy-compact, thus, we further decorrelate feature maps using the PCA. Examples of generated feature maps and the rotated feature maps by the PCA are shown in Figure~\ref{fig:features}. It is observed that more zeros are generated in the bottom-right corner and large values are centered in the top-left corner in the rotated feature maps, which benefits the following entropy coder to reduce the rate. The CAE-based method outperforms JPEG and achieves a 13.7\% BD-rate decrement on the Kodak database images compared to JPEG2000. The detailed discussions refer to the paper~\cite{IEEEexample:PCS}.

\subsection{Discussion on GAN-based Compression}

We conduct some experiments on the \emph{CLIC} validation dataset to discuss the performance. First, the effect of input sizes for one image is listed in Table~\ref{tab.inputsize}, where $64\times64\times3\rightarrow1024$ denotes that the input size is $64\times64\times3$ and the code size $N$ is 1024. It is observed that the input size $128\times128$ obtains the best PSNR because the tested image size is around 1080p, resulting that $128\times128$ is a proper size for semantics reconstruction of GANs. Second, the effect of code sizes and interpolation sizes is given in Table~\ref{tab.codesize}. Code size is the length of generated compressed code $N$. We set the input size as $128\times128$. Along with the increase of code sizes, PSNR and MS-SSIM increases. The GAN with fixed code size cannot provide good performance for all the images with different textures, so an adaptively switchable encoder for GANs with different code sizes will be studied in the future. To obtain variable bit rates, we add one bicubic interpolation filter with different scales as the preprocessing. From Table~\ref{tab.codesize}, by interpolating the size from (W, H) to (2W, 2H), PSNR increases by about 2.2dB, MS-SSIM increases by around 0.10. Meanwhile, the rate increases up to almost 4 times. The effect of different quantization bits is shown in Table~\ref{tab.quanbit}. We set the code size as 256 with (4W, 4H) interpolation. Too few bits, e.g. 5 bit, will destroy the reconstruction quality significantly. Similar to CAE-based method, we also apply PCA to further reduce the code size. For the \emph{CLIC} challenge, the entry \emph{Gcode} uses the architecture of $128\times128\times3\rightarrow128$ with (3W, 3H) interpolation, 8-bit quantization and PCA rotation.

\begin{table}
\begin{center}
\begin{footnotesize}
\begin{tabular}{|p{74pt}|l|l|p{35pt}|}
\hline
\textbf{Input size} & \textbf{PSNR(dB)}  & \textbf{MS-SSIM} & \textbf{Rate(bpp)} \\
\hline
$64\times64\times3\rightarrow1024$  &22.73  &0.745 &0.781\\
$128\times128\times3\rightarrow1024$ &23.95   &0.897 &0.225 \\
$256 \times 256\times3\rightarrow1024$ &17.18 &0.699 &0.050 \\
\hline
\end{tabular}
\end{footnotesize}
\end{center}
\caption{The effect of different input sizes.}
\label{tab.inputsize}
\end{table}

\begin{table}
\begin{center}
\begin{footnotesize}
\begin{tabular}{|p{33pt}|p{40pt}|p{32pt}|l|p{35pt}|}
\hline
\textbf{Code size} & \textbf{Interp. Size} & \textbf{PSNR(dB)}  & \textbf{MS-SSIM} & \textbf{Rate(bpp)} \\
\hline
$64$  &(W,H) &22.195  &0.753 &0.024\\
      &(2W,2H) &24.213  &0.856 &0.086\\
      &(4W,4H) &26.451  &0.928 &0.329\\
\hline
$128$ &(W,H) &23.126   &0.791 &0.042 \\
      &(2W,2H) &25.335  &0.901  &0.162\\
      &(4W,4H) &27.801  &0.941  &0.389\\
\hline
$256$ &(W,H)   &23.962  &0.831  &0.071 \\
      &(2W,2H) &26.308  &0.924  &0.274 \\
      &(4W,4H) &29.262  &0.957  &0.792 \\
\hline
$1024$ &(W,H)  &25.121 & 0.896 &0.261 \\
      &(2W,2H) &27.474 &0.947 &0.981 \\
\hline
\end{tabular}
\end{footnotesize}
\end{center}
\caption{The effect of different code sizes and interpolation sizes.}
\label{tab.codesize}
\end{table}

\begin{table}
\begin{center}
\begin{footnotesize}
\begin{tabular}{|l|l|l|l|}
\hline
\textbf{Quan. bit} & \textbf{PSNR (dB)}  & \textbf{MS-SSIM} & \textbf{Rate (bpp)} \\
\hline
8 bit &27.932  &0.952 &0.764\\
7 bit &27.788  &0.942 &0.472 \\
6 bit &27.313  &0.901 &0.352 \\
5 bit &25.179  &0.784 &0.233 \\
\hline
\end{tabular}
\end{footnotesize}
\end{center}
\caption{The effect of different quantization bits.}
\label{tab.quanbit}
\end{table}

\subsection{Comparison Results}

\begin{figure}[tb]
\centering
\subfigure{
\label{Fig.sub.a1}
\includegraphics[width=0.235\textwidth]{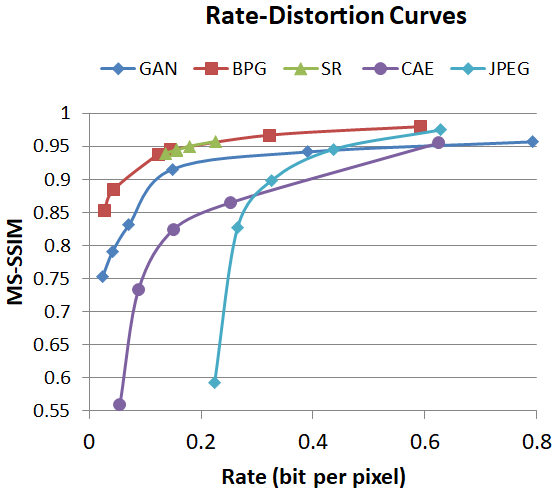}}
\subfigure{
\label{Fig.sub.a2}
\includegraphics[width=0.226\textwidth]{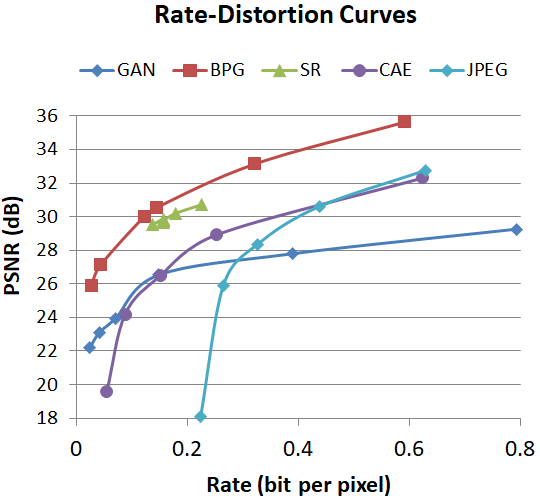}}
\caption{RD curves of three methods.}
\label{fig:RDcurves}
\end{figure}

In this section, we use the \emph{CLIC} validation dataset for a fair evaluation. The RD curves with MS-SSIM and PSNR are shown in Figure~\ref{fig:RDcurves}. RD curves for super-resolution is short because it is conducted by changing the threshold in the adaptive strategy with the fixed quantization parameter (QP) value in BPG codec. By changing the QP, super-resolution can also achieve a wide range of RD curves. Several observations are summarized from RD curves. 1) CAEs are better than JPEG in case of lossy compression due to the inherent property of autoencoder. Autoencoders can reduce the dimension to extract the compressed presentation from images, so CAEs outperform JPEG and JPEG2000. 2) GANs perform better with low bit rate than that with high bit rate, so GANs tend to achieve large compression ratio. Meanwhile, GANs have better performance on MS-SSIM than PSNR, because the reconstruction of GANs is based on the distribution of the image data, which is friendly to human visual system. Especially for MS-SSIM, GANs have stable performance from 0.2bpp to 0.8bpp. 3) SR achieves the best performance among these three methods, because it takes the advantages of both emerging algorithms BPG and machine learning based super-resolution filters. Promising results can be expected to outperform BPG by adding better super-resolution filters, if more computational resources can be provided.


The comparison for three methods with the rate constraint of 0.15bpp is shown in Table~\ref{tab.comp}. It is observed that SR-based method is quite close to BPG. GAN and CAE based architectures are better than JPEG. Especially, GANs and CAEs have the similar PSNR, but GANs are much better than CAEs in terms of relatively subjective MS-SSIM.

\begin{table}
\begin{center}
\begin{footnotesize}
\begin{tabular}{|l|l|l|l|}
\hline
\textbf{Codecs} & \textbf{PSNR (dB)}   & \textbf{MS-SSIM} & \textbf{Rate (bpp)} \\
\hline
JPEG &25.82 &0.853 &0.133 \\
CAE &26.48  &0.825 &0.151 \\
GAN & 26.53  &0.915 & 0.148\\
SR  &30.00 &0.947 &0.143 \\
BPG &30.85 &0.948 &0.149
 \\
\hline
\end{tabular}
\end{footnotesize}
\end{center}
\caption{Performance comparison with 0.15bpp constraint.}
\label{tab.comp}
\end{table}

\vspace{-2mm}
\section{Conclusion and Future Work}

End-to-end deep learning based compression is a challenging work. In this paper, we propose three architectures using CAEs, GANs and SR, for compression, and discuss their performance. Results demonstrate that 1) CAEs are better than traditional transforms for lossy compression, and are expected to be used as a feature extractor. 2) GANs show potential advantages on large compression ratio and subjective quality reconstruction. 3) SR-based compression achieves the best coding performance among them. In the future work, we will design a target-adaptive encoders by switching multiple GANs with variable code sizes. We will add mean opinion scores (MOS) evaluations to illustrate the effectiveness. Combining individual advantages of each method will contribute to a better compression algorithm.

{\small

}

\end{document}